\documentclass[12pt]{revtex4-1}

\usepackage{palatino}
\usepackage{graphicx}
\usepackage{amsmath}
\usepackage{longtable}
\usepackage[welsh,english]{babel}
\usepackage{tabularx}

\begin{document}

\title{The optimal division between sample and background measurement time for photon counting experiments}
\author{Brian Richard Pauw}
\email{brian@stack.nl}
\affiliation{--}
\author{Samuel Tardif}
\email{samuel.tardif@gmail.com}
\affiliation{--}
\date{\today}

\maketitle

\section{Outline}
\emph{Note by the author: It was found after this derivation, that the work presented here had been derived differently but with similar conclusions by Steinhart and Plestil \cite{Steinhart-1993}. We nevertheless believe that this simplified derivation may be of use to some readers.} 

Usually, equal time is given to measuring the background and the sample, or even a longer background measurement is taken as it has so few counts. While this seems the right thing to do, the relative error after background subtraction improves when more time is spent counting the measurement with the highest amount of scattering. As the available measurement time is always limited, a good division must be found between measuring the background and sample, so that the uncertainty of the background-subtracted intensity is as low as possible.

Herein outlined is the method to determine how best to divide measurement time between a sample and the background, in order to minimize the relative uncertainty. Also given is the relative reduction in uncertainty to be gained from the considered division. It is particularly useful in the case of scanning diffractometers, including the likes of Bonse-Hart cameras, where the measurement time division for each point can be optimized depending on the signal-to-noise ratio. 

The optimum setting for machines with photon-counting two-dimensional detectors has to be further evaluated, but the intention is to include that in this note.

\section{The calculation}
We assume that the number of background intensity photons $I_b$, measured for a time $t_{b}$ is subtracted from the sample measurement photon count $I_s$ which was measured for a time $t_s$, to result in the background subtracted count rate $C_{bs}$:
\begin{equation}\label{eq:sIbgs}
C_{bs}=\frac{I_s}{t_s}-\frac{I_b}{t_b}
\end{equation}
Defining the sample uncertainty as $\Delta I_s=\sqrt{I_s}$ and the background uncertainty similarly as $\Delta I_b=\sqrt{I_b}$, the uncertainty $\Delta C_{bs}$ would then be:
\begin{equation}\label{eq:sIerr}
\Delta C_{bs}=\sqrt{\left(\frac{\Delta I_s}{t_s}\right)^2+\left(\frac{\Delta I_b}{t_b}\right)^2 }=\sqrt{\frac{I_s}{t_s^2}+\frac{I_b}{t_b^2}}
\end{equation}
Defining the number of counted photons $I$ to be a multiplication of the count rate $C$ and the measurement time, we get $I_b=C_b t_b$ and $I_s=C_s t_s$. Further defining the signal-to-noise ratio $g=\frac{C_s}{C_b}$, the total time $t_t=t_b+t_s$ and the fraction of time spent measuring the sample $f=\frac{t_s}{t_t}$, we can define our relative uncertainty in terms of signal-to-noise ratio and time fraction:

\begin{equation}\label{eq:sErel}
\frac{\Delta C_{bs}}{C_{bs}}=\sqrt{\frac{\frac{C_s}{t_s}+\frac{C_b}{t_b}}{(C_s-C_b)^2}}=\sqrt{\frac{1}{C_b t_t}}\sqrt{\frac{\frac{g}{f}+\frac{1}{(1-f)}}{(g-1)^2}}
\end{equation}

We then can try to find the optimum by locating the value for $f$ where the derivative of equation \ref{eq:sErel2} is zero:

\begin{equation}\label{eq:sd1}
\frac{\partial \frac{\Delta C_{bs}}{C_{bs}}}{\partial f}=\frac{\partial}{\partial f}\sqrt{\frac{1}{C_b t_t}}\sqrt{\frac{\frac{g}{f}+\frac{1}{(1-f)}}{(g-1)^2}}=0
\end{equation}
\begin{equation}\label{eq:sd2}
0=\frac{\partial}{\partial f}\sqrt{\frac{\frac{g}{f}+\frac{1}{(1-f)}}{(g-1)^2}}
\end{equation}
which, given $0<f<1$, is true for
\begin{equation}\label{eq:finalf}
f=\frac{g-\sqrt{g}}{g-1}
\end{equation}

We can calculate the relative reduction in uncertainty compared to the 50/50 case (i.e. equal time spent on background and sample measurements) as:
\begin{equation}\label{eq:finalopt}
\frac{\frac{\Delta C_{bs}}{C_{bs}}\mid_\mathrm{50/50}-\frac{\Delta C_{bs}}{C_{bs}}\mid_\mathrm{optimal}}{\frac{\Delta C_{bs}}{C_{bs}}\mid_\mathrm{50/50}}=\frac{\sqrt{\frac{2g+2}{(g-1)^2}} - \sqrt{ \frac{ \frac{g^2-g}{g-\sqrt{g}}+\frac{g-1}{\sqrt{g}-1}}{(g-1)^2}} }{\sqrt{\frac{2g+2}{(g-1)^2}}}
\end{equation}

\begin{figure}
   \centering
   \includegraphics[angle=0, width=0.75\textwidth]{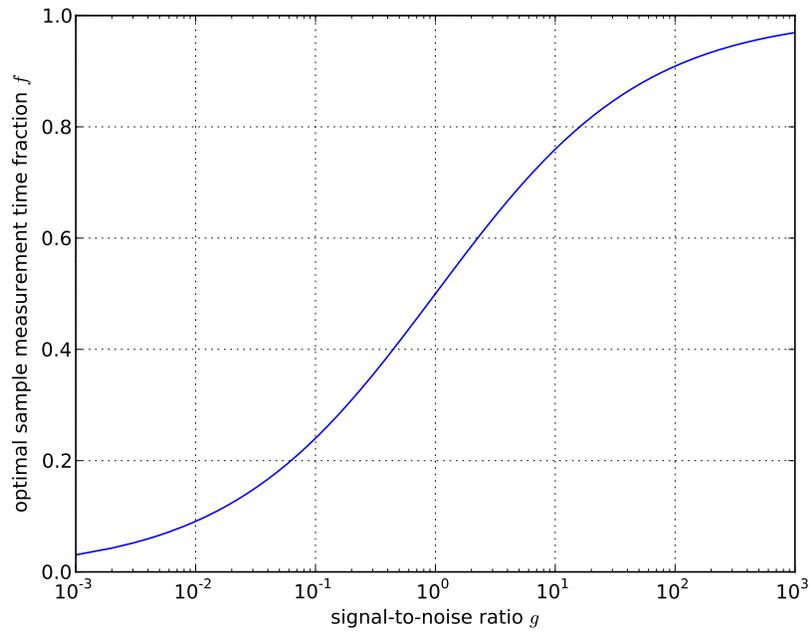} 
   \caption{The optimal fraction of time $f$ spent measuring the sample as opposed to measuring the background, as a function of the signal-to-noise ratio $g$.}
    \label{fg:finalf}
\end{figure}
\begin{figure}
   \centering
   \includegraphics[angle=0, width=0.75\textwidth]{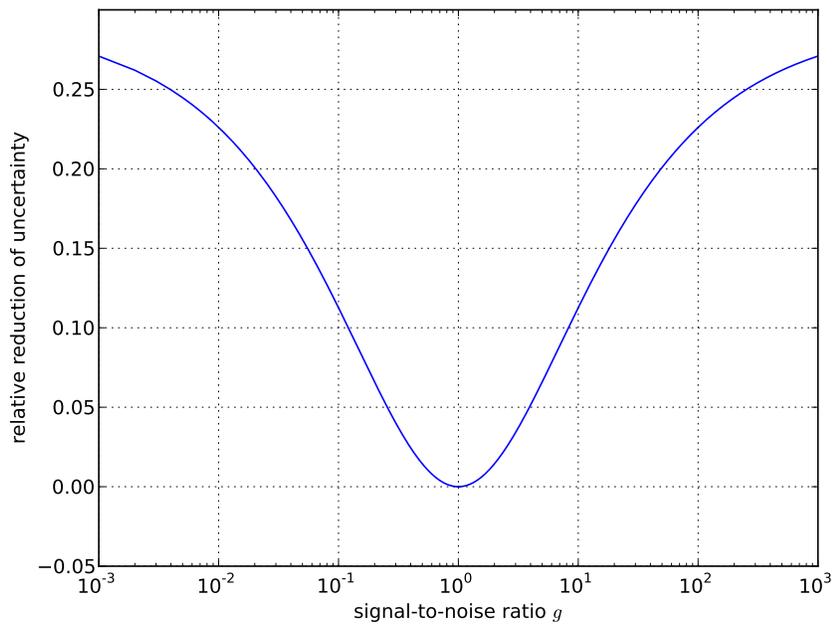} 
   \caption{The reduction in error that can be obtained by dividing the time optimally between sample and background measurement, as a function of the signal-to-noise ratio $g$.}
    \label{fg:finalopt}
\end{figure}

\section{Conclusions}
Figures \ref{fg:finalf} and \ref{fg:finalopt} show the optimal division of time between sample and background, and the reduction in uncertainty obtained through this optimization, respectively. These figures clarify that the reduction in uncertainty may be worth the trouble of a quick determination of the signal-to-noise ratio, especially in areas where this ratio strongly deviates from unity.

A quick scan of sample and background may be used to automatically select the most optimal use of measurement time, in particular for scanning (small- and wide-angle) diffractometers (including oddities such as Bonse-Hart cameras) where the measurement time \emph{per point} can be freely tuned.

\bibliography{/Users/brian/Documents/bibliography}

 \end{document}